\documentclass[twocolumn,superscriptaddress,pre,longbibliography]{revtex4-1}

\usepackage[utf8]{inputenc} \usepackage{color}
\usepackage[T1]{fontenc}
\usepackage{graphicx}
\usepackage{amsmath}
\usepackage{amssymb}
\usepackage{xspace}
\usepackage{hyperref}

\begin{document}
\title{Phase transition in the computational complexity of the\\ shortest common
superstring and genome assembly}

\author{L. A. Fernandez}\affiliation{Departamento de Física Teórica, Universidad Complutense, 28040 Madrid, Spain}\affiliation{Instituto de Biocomputación y Física de Sistemas Complejos (BIFI), 50018 Zaragoza, Spain}
\author{V. Martin-Mayor}\affiliation{Departamento de Física Teórica, Universidad Complutense, 28040 Madrid, Spain}\affiliation{Instituto de Biocomputación y Física de Sistemas Complejos (BIFI), 50018 Zaragoza, Spain}
\author{D. Yllanes}\affiliation{Chan Zuckerberg Biohub, 499 Illinois Street, San Francisco, CA 94158, USA} \affiliation{Instituto de Biocomputación y Física de Sistemas Complejos (BIFI), 50018 Zaragoza, Spain}

\newcommand{\changes}[1]{#1}
\newcommand{\changesbis}[1]{{#1}}
\newcommand{\NF}{\ensuremath{N_{\text{frag}}}\xspace}
\newcommand{\lf}{\ensuremath{\ell_{\text{frag}}}\xspace}
\newcommand{\lo}{\ensuremath{\ell_{\text{ordered}}}\xspace}
\newcommand{\ps}{\ensuremath{p_{\text{success}}}\xspace}
\newcommand{\xc}{\ensuremath{x_{\text{c}}}\xspace}
\newcommand{\dm}{\ensuremath{d_{\text{max}}}\xspace}

\begin{abstract} 
Genome assembly, the process of reconstructing a long genetic sequence by
aligning and merging short fragments, or reads, is known to be NP-hard,
either as a version of the shortest common superstring problem or in a 
Hamiltonian-cycle formulation. That is, the computing time is 
believed to grow exponentially with the the problem size in the worst
case. Despite this fact, high-throughput technologies and
modern algorithms currently allow bioinformaticians to 
\changes{handle} datasets of billions of reads. Using methods from statistical mechanics, we address
this conundrum by demonstrating the existence of a phase transition in 
the computational complexity of the problem and showing 
that practical instances always fall in the `easy' phase (solvable 
by polynomial-time algorithms). In addition, we propose a Markov-chain 
Monte Carlo method that outperforms common deterministic algorithms 
in the hard regime.
\end{abstract}

\maketitle 

\section{Introduction}
Sequence assembly is one of the fundamental problems in 
bioinformatics. Since an organism's whole genetic material cannot be read 
in one go, current technologies build on strategies
where the genome (or a portion of it, such as a chromosome)
is randomly fragmented in shorter reads,  which then have to be ordered and merged
to reconstruct the original sequence. In a naive formulation, one
would look for the shortest sequence that contains all of the individual
reads, or Shortest Common Superstring (SCS). This is, in principle,
a formidable task, since the SCS belongs to the so-called 
NP-complete class of problems~\cite{maier:78,gallant:80}, for which no efficient algorithms are 
known (or believed) to exist. More precisely, NP denotes
a large family of problems that are verifiable in polynomial time, 
meaning that potential solutions can be checked in a time that grows 
at most as a power of the size of the input. A problem is then 
termed NP-complete if it is in NP and at least as hard as any problem in NP. This is 
a relevant notion because it is believed, but not proven, that not all 
NP-complete tasks can be solved in polynomial time. See, \emph{e.g.},
\cite{knuth:74,papadimitriou:94,cormen:09} for more precise definitions and
examples.

As hinted above, the formulation of genome assembly as an SCS problem
is not quite correct. This is because our assumption 
of parsimony is not true: most genomes contain \emph{repeats}, multiple
identical stretches of DNA, which the SCS would collapse. A 
formulation of the assembly problem that takes this issue into account
can be made using de Bruijn graphs~\cite{pevzner:01}, but the task
can still be proven to be NP-hard by reduction from SCS~\cite{medvedev:07}.
Alternative approaches to assembly have been proposed, for instance the 
string-graph representation~\cite{myers:05}, but this model has also
been shown to be NP-hard, by reduction from the Hamiltonian-cycle problem~\cite{medvedev:07}.

In short, no polynomial-time algorithms are known (or even believed)
to exist to solve the sequence-assembly problem in its general
formulation. Despite this fact, with current high-throughput
sequencing techniques and assembly algorithms, datasets of billions of
reads are regularly assembled \changes{(at least at the contig
  level)}~\cite{schatz:10, compeau:11,
  miller:10,zhang:11,el-metwally:13,levy:16}. This achievement is in
stark contrast with the state of the art for the travelling salesman,
a closely related NP-complete problem, for which managing as few as
$10^4$ `cities' is exceedingly difficult and the largest instance
solved to date featured $~120\,000$
locations~\cite{applegate:06,tsp:waterloo}.

The way out of this apparent contradiction is the general notion that, while
in the worst case an NP-complete problem takes exponential time
to solve, typical instances might be much easier. This observation could explain
the, at least apparent, success of heuristic methods but it needs to be formalised:
what is a `typical' instance and how likely is the worst-case scenario? 
As a path towards answering those questions, it has been
observed that in several problems in computational biology, small ranges
of parameters cover all the interesting cases. The question is, then, whether we can
identify the right variables and whether the problem can be solved in polynomial
time for fixed values of these parameters~\cite{bodlaender:95}. This parametric
complexity paradigm has been applied to genome assembly, either using statistical
analyses or analytical methods, suggesting that, in some relevant limits, the problem
can indeed be solved correctly with polynomial algorithms~\cite{nagarajan:09, kingsford:10,bresler:13}.

In this work we present an alternative approach to this question,
based on the methods of statistical mechanics. The applicability of
the models of statistical physics to the issue of NP-completeness has
been conjectured since the 1980s~\cite{fu:85,mezard:87} and is now
well understood~\cite{mezard:09}. More to the point, 
phase diagrams for complexity can be defined for some problems.
For instance, in a seminal work~\cite{monasson:99} 
all the constraints in the problem
can be satisfied only when a certain parameter is smaller than its critical
value, and the computationally hard problems arise only in the
neighbourhood of the phase boundary.

\changes{We show that a similar result can be obtained for the
  SCS. Yet, we depart from the paradigm of Ref.~\cite{monasson:99} in
  the sense that computationally hard problems become the rule, rather
  than the exception, in one of our two phases, not just at the
  critical point.

  It is worth emphasizing that, unlike assembly, the SCS
  problem \emph{always} has a well-defined solution\changesbis{, which might not be unique} (unfortunately, in
  some regions of parameter space this solution is very hard to
  find). Instead, the assembly problem is well posed only if the whole
  genome is covered. It is a classic result~\cite{lander:88} that, if
  $L$ is the length of our (portion of) genome and \lf\ the length of
  the reads, in order to ensure that the whole genome is covered with
  probability $1-\epsilon$, the number \NF\ of reads must satisfy
  $\NF = (L/\lf) \log (\NF/\epsilon)$. Therefore, in practical
  applications one is interested in oversampling the genome (the
  oversampling ratio is termed coverage). Our main result in this
  respect is that the regime of full coverage corresponds precisely to
  the easily solvable phase of the SCS problem. Therefore, whenever
  the assembly problem is well posed, the corresponding solution can
  be found in polynomial time.

  A final note of warning is in order: we shall always use assembly language
(genomes and reads, rather than superstrings and strings) in order to unify the
nomenclature.

  The remaining part of this paper is organized as follows. In
  Sect.~\ref{sect:def} we define the two problems considered here, the
  SCS and the assembly. In Sect.~\ref{sec:data} we discuss our data
  collection and computational approaches. The analysis of the phase
  transition is presented in Sect.~\ref{sect:transition}. An
  alternative algorithm for the hard phase of the SCS is presented in
  Sect.~\ref{sec:swap}. Our conclusions are given in
  Sect.~\ref{sect:conclusion}. We provide additional details on the
  employed algorithms in the appendices.
  
}

\section{The SCS and sequence assembly}\label{sect:def}
As explained in the introduction, there are two different, yet related problems: \begin{itemize}
\item \emph{Shortest Common Superstring (SCS).\/} Given \NF\ sequences of \lf\
letters taken from a common alphabet, find the shortest sequence of letters
that contains every one of the $\NF$ fragments.  \item \emph{Ex novo genome
reconstruction.\/} Read \NF\ fragments randomly chopped from a piece of genome.
For simplicity, we shall assume that each fragment contains the same number of
letters \lf.  Our problem is reconstructing the original genome from these
reads.
\end{itemize}
Under favourable circumstances on the ensemble of reads, the solution
of the SCS problem is also the solution of the assembly problem.  Our
main emphasis will be in the combinatorial optimisation problem,
namely the SCS. Reading errors are a real complication in assembly,
but effective methods are known to handle them. Since errors do not
add to the exponential (in genome size) hardness of the problem, we
shall ignore them.

We study the SCS in its formulation as an asymmetric
travelling-salesman problem, where one tries to find the permutation
of fragments that has the maximum overlap between consecutive segments
and, therefore, the minimal total length of the resulting superstring
once overlapping segments are collapsed.  For instance, the SCS of the
strings TTGAA, AGTTG is AGTTGAA.  Our reads are taken from a circular
genome of length $L$ base pairs (we use the natural four-letter
alphabet: A, C, G, T).  For our main study we choose all bases in the alphabet randomly (independent picks with uniform
probability), but we have also checked that our main results extend to a natural
genome, namely that of the swinepox virus.
\begin{figure}[t]
\includegraphics[width=\linewidth]{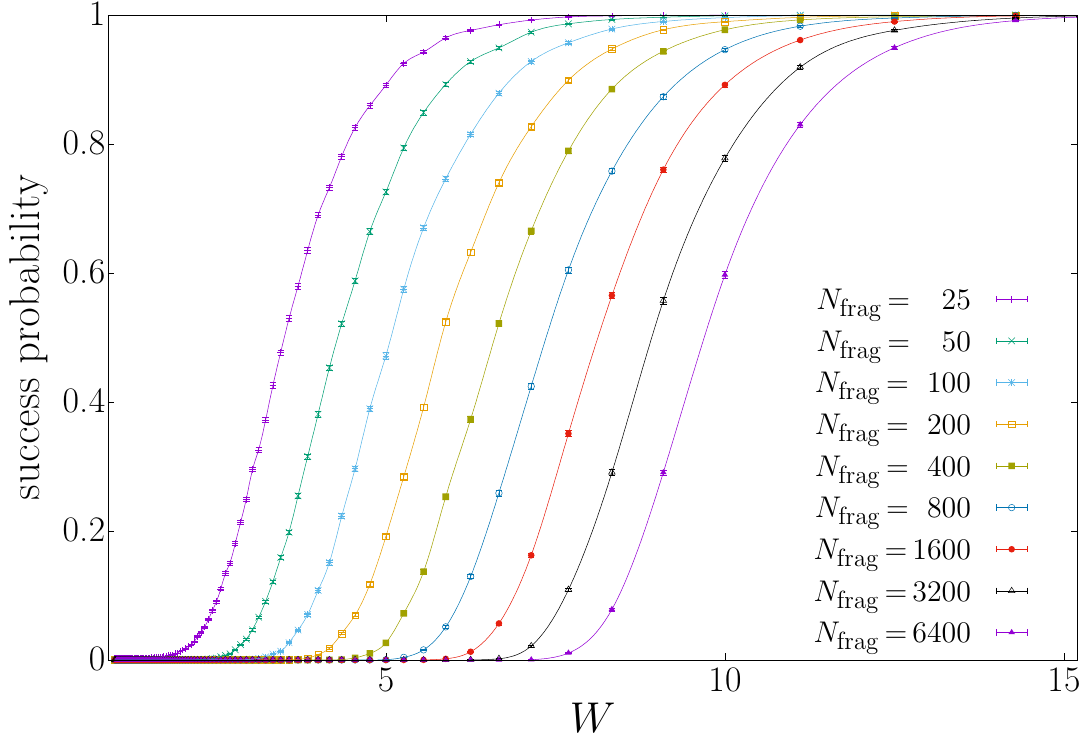}
\includegraphics[width=\linewidth]{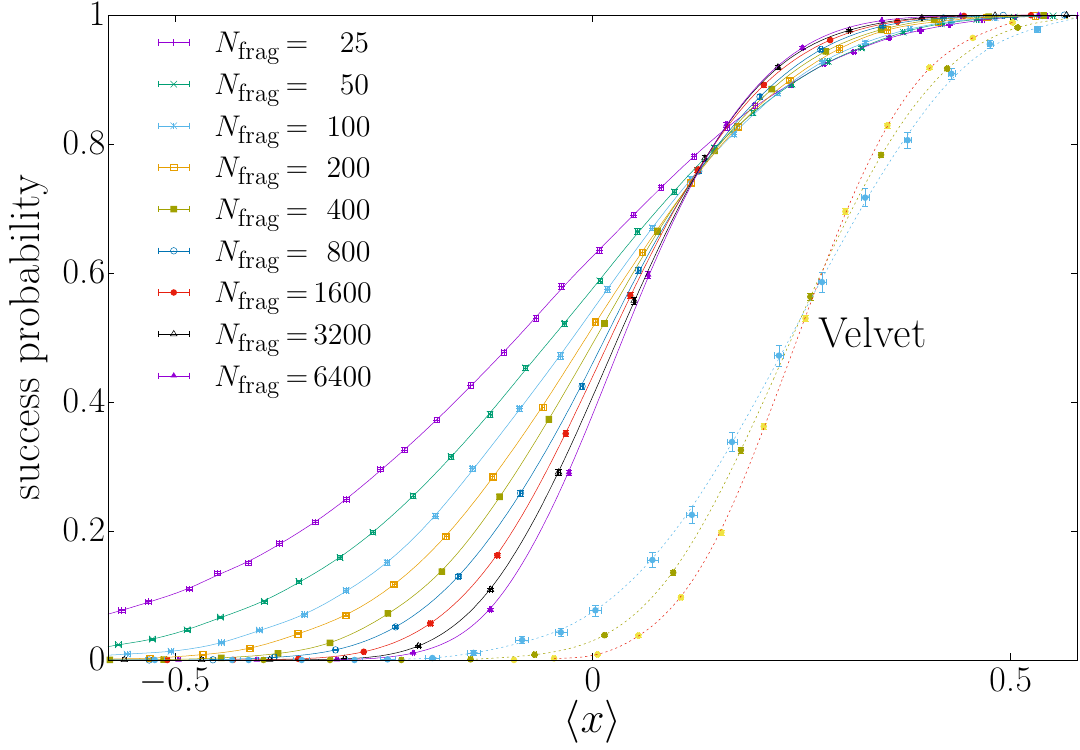}
\caption{\textbf{Performance of common algorithms for the shortest common
superstring problem.} \emph{Top:} Probability of finding a successful solution
(see text) using a greedy algorithm as a function of the coverage $W$,
eq.~(\ref{eq:W}), for several values of the number of fragments (reads) \NF\ and
for fragment length $\lf=100$. For large coverage values, the algorithm always
succeeds.  \emph{Bottom:} In terms of the correct scaling variable $x$,
eq.~(\ref{eq:x}), based on the ratio between the average maximum distance
between fragments and \lf, the \ps\ curves for different \NF\ cross, which we
interpret as the onset of a phase transition at some critical \xc.  The value
of \xc\ is algorithm dependent, but the qualitative behaviour is the same for
more sophisticated methods.  As a demonstration, we also show the results using
\emph{Velvet}, which employs an algorithm based on de Bruijn
graphs.   \label{fig:gloton-velvet}} \end{figure}

A naive approach to an assembly problem emphasises the covering fraction:
\begin{equation}\label{eq:W}
  W= \frac{\NF\lf}{L}\,.
\end{equation} 
$W<1$ implies that the SCS is shorter than the genome \changes{(obviusly, succesful assembly is impossible under these circumstances)}. 
In typical instances of assembly
$W\gg 1$  ($W\sim1000$ is not uncommon
with high-throughput techniques).

Given a set of reads, obtaining the SCS is NP-hard. Since we are
taking our fragments from a known long string, however, we always have
a candidate solution, \changes{as we now explain. Since the genome is
  known to us beforehand, we can exactly locate the position in the
  genome of every fragment. If we order these starting points in
  increasing order, we obtain by merging overlaps a candidate solution
  for the SCS. We name $\lo$ the length of the candidate solution,
  which often turns out to be the exact solution ($\lo=L$) when
  $W\gg 1$. This is the common situation in applications. Yet, when
  $W<1$, $\lo$ is guaranteed to be smaller than $L$. Furthermore, the
  ordered sequence may actually be a very bad solution for SCS
  problem, when $W\ll 1$. Indeed, in the limit $W\to 0$, nothing
  distinghuises the ordered solution from a random ordering. In
  the intermediate regime, $W\sim 1$, the ordered sequence is a good
  guess for the SCS (and for assembly).}

For a given algorithm, a run whose resulting
superstring length $\ell$ is $\ell\leq \lo$ will be considered
successful. In practice, in the $W\gg1$
region, one never finds $\ell<\lo=L$ and the original genome coincides with the
SCS.

\changes{\section{Creating and sampling the data set}\label{sec:data}}

The main classifying feature for our simulations is the number of
fragments, $\NF$. For every $\NF$ we create a set of $N_{\text{chro}}$
synthetic circular chromosomes. As discussed above, we have considered
both random chromosomes ---in which each letter is extracted with
uniform probability randomly from the four-letter alphabet--- or
extracted reads from the genome of the swinepox virus (downloaded from
GenBank, accession number NC\_003389).

\changes{To generate the fragments, we proceed as follows. We independently and
randomly generate \NF integers uniformly distributed in
$\{1,2,\ldots,L\}$ ($L$ is the length of the \emph{circular}
chromosome). Each integer is regarded as the starting point of a
fragment of length \lf.}

\changes{We have analyzed our data using two algorithms: a
  version of the greedy algorithm, which we name \emph{Glotón}, and
  \emph{Velvet}~\cite{zerbino:08}, a commonly used program for genome
  assembly based on de Bruijn graphs (see Appendix~\ref{sec:gloton}
  for a description of the \emph{Glotón} algorithm and
  Appendix~\ref{sec:velvet} for details on our simulation parameters
  with \emph{Velvet}). It will be important that \emph{Glotón} has a
  stochastic component, while \emph{Velvet} is deterministic}.

\changes{
We attempt to reconstruct the chromosome from this set of reads
$N_{\text{attempts}}$ times (we generate just one set of reads for
each chromosome; we set $N_{\text{attempts}}=1$ for \emph{Velvet}).
The success probability for a given chromosome is the fraction of
the $N_{\text{attempts}}$ assembly attempts that meet our success
criterion $\ell \leq \ell_\text{ordered}$. Specifically, let $\ell_{i,j}$ be
the length obtained on the $j$-th attempt for the $i$-th chromosome. We have
for \emph{Glotón}
\begin{equation}\label{eq:def-pi}
  p_{\text{success}}^{(i)}=\frac{1}{N_{\text{attempts}}}\sum_{j=1}^{N_{\text{attempts}}}\boldsymbol{1}(\ell_{i,j}\leq\ell_\text{ordered})\,,
\end{equation}
where $\boldsymbol{1}$ is the indicator function. For the deterministic Velvet, we adopt
a sliglty different definition, see Appendix~\ref{sec:velvet}, such
that $p_{\text{success}}^{(i)}=0,1$. From now on, we shall refer to
$p_{\text{success}}^{(i)}$ as the individual success probability.

The total success
probability is  just the average over the $N_{\text{chro}}$ chromosomes
of the individual success rates
\begin{equation}\label{eq:def-ps}
  \ps=\frac{1}{N_{\text{chro}}}\sum_{i=1}^{N_{\text{chro}}}p_{\text{success}}^{(i)}\,.
\end{equation}}
We compute in a similar way the
variance ---or covariance--- of the individual success probabilities.

We have set $N_{\text{attempts}}=100$ for 
the \emph{Glotón\/} and \emph{segment-swap\/} algorithms (described
below) and, as we said above, $N_\text{attempts}=1$ for \emph{Velvet}. We use
$N_{\text{chro}}=10000$ for \emph{Glotón\/} (the only exception is in
Figure~\ref{fig:scaling}, where $N_{\text{chro}}=100000$ for
$\NF\leq 800$). On the other hand, we have contented ourselves with
$N_{\text{chro}}=1000$ for the costlier \emph{Velvet\/} and
\emph{segment-swap\/} algorithms (in Figure~\ref{fig:gloton-T0}, for $\NF=800$
and \emph{segment-swap}, we use $N_{\text{chro}}=100$).\\

\section{The success probability and a phase transition in the complexity}\label{sect:transition}
We want to characterise the hardness of the problem in terms of the
success probability \ps for a run of a simple algorithm (\emph{i.e.},
one that ends in polynomial time). Here we consider two, namely
\emph{Glotón}, and \emph{Velvet}.

It is our goal to understand quantitatively the behaviour of \ps\ as a function
of $L$, \lf\ and \NF. Ideally, one would be able to simplify the three-variable
function $\ps=f(L,\NF,\lf)$ into a function of a single \emph{scaling} variable
$x$. A first attempt, shown in Figure~\ref{fig:gloton-velvet}-top, plots \ps\
as a function of $W$ for $\lf=100$ and various values of \NF\ for the
\emph{Glotón} (see Appendix~\ref{sec:data} for more
details on these simulations).  We can see two regimes: for large $W$, this algorithm always
succeeds, while for small $W$ it always fails. Recall that, in the large-$W$
regime, success effectively means reconstructing the original genome, while in
the small-$W$ regime it means finding a good approximation to the SCS given by
\lo.

Comparing the different curves in Figure~\ref{fig:gloton-velvet}-top,
we see that $W$ is not a good scaling variable, since a
clear dependence on \NF\ remains. A more natural candidate is the maximum
distance \dm\ between the starting points of reads consecutive in the
original genome. Notice that the genome is fully covered by the
\NF\ fragments if and only if $\dm \leq \lf$. For a given realisation
of the problem, we define \begin{equation}\label{eq:x} x = 1 -
  \dm/\lf.  \end{equation} We can also define $\langle x\rangle$ as
the ensemble average of $x$ for all possible genomes of length $L$ and
all possible choppings with \NF\ and \lf. It can be shown that
\changes{\begin{equation}\label{eq:discretization}
  \langle d_\text{max}\rangle \sim \log\NF/W,
\end{equation}}for instance by discretization of
the continuum calculation in~\cite{schlemm:14}. In each 
of the different curves of Figure~\ref{fig:gloton-velvet}
\NF\ and \lf\ are fixed and $\langle x\rangle$ (or $W$) are 
varied by changing $L$.

Plotting \ps\ as a function of $\langle x\rangle$, we can see that the curves
for different \NF\ cross at some \xc, while their shape approaches a step
function as \NF\ increases. This is the characteristic finite-size scaling
behaviour at a phase transition~\cite{amit:05}. Interestingly enough, the
\emph{Glotón} and \emph{Velvet} algorithms have the same behaviour, just with a
shift in \xc.
Again, this is typical of phase transitions, where the critical
point depends on the details of the model, but the scaling behaviour is
universal.  We thus propose that, in the large-\NF\ limit, the SCS
problem experiences a phase transition that separates a large-$x$
where polynomial-time algorithms always succeed from a small-$x$ regime where
they always fail. All practical applications of genome
assembly are in the (very) large-$x$ regime, which explains how it is routinely
possible to solve problems that are, in principle, NP-hard with millions of
fragments.

\begin{figure}[t] \includegraphics[width=\linewidth]{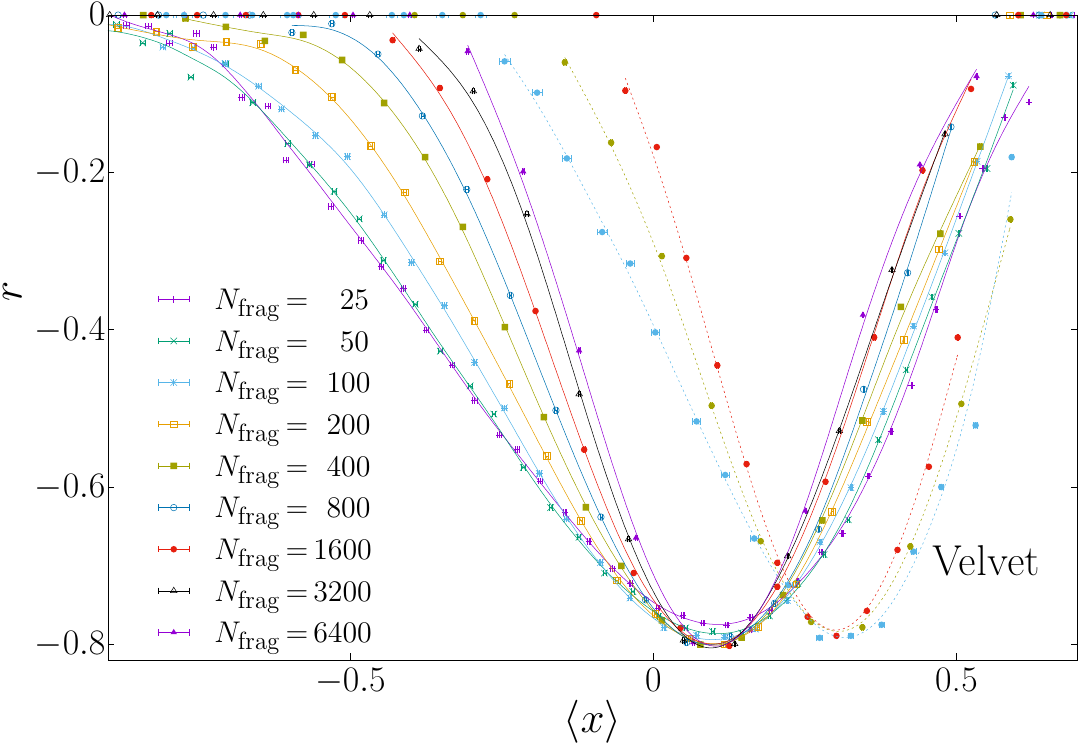}
\caption{\textbf{Location of the critical point.} The critical point
can be determined by looking for the value of $\langle x\rangle$ where 
fluctuations are largest. We plot the correlation coefficient
between the scaling variable $x$ and the success probability 
for single realisations of the \NF\ reads. The absolute value of $r$
has a maximum at the critical point
\xc.  \label{fig:gloton-r}} \end{figure}

We can study the critical regime more quantitatively by looking not just at
means but at fluctuations. In particular, we plot in Figure~\ref{fig:gloton-r}
the correlation coefficient $r$ between $x$, as computed for a particular set
of fragments, and \changes{the individual success probability~\eqref{eq:def-pi} of a polynomial
algorithm for that particular set}. Away from the critical point, $r$ is very
small but in the critical regime a strong (anti)correlation is observed. In fact, the
minimum of $r$ as a function of $\langle x\rangle$ is probably the best way of
locating \xc. Notice that \xc\ is \changes{lower} for \emph{Glotón}
than for \emph{Velvet}, which is perhaps unsurprising, since the latter was not
designed as an SCS solver (see note in Appendix~\ref{sec:velvet}).

\begin{figure} \includegraphics[width=\linewidth]{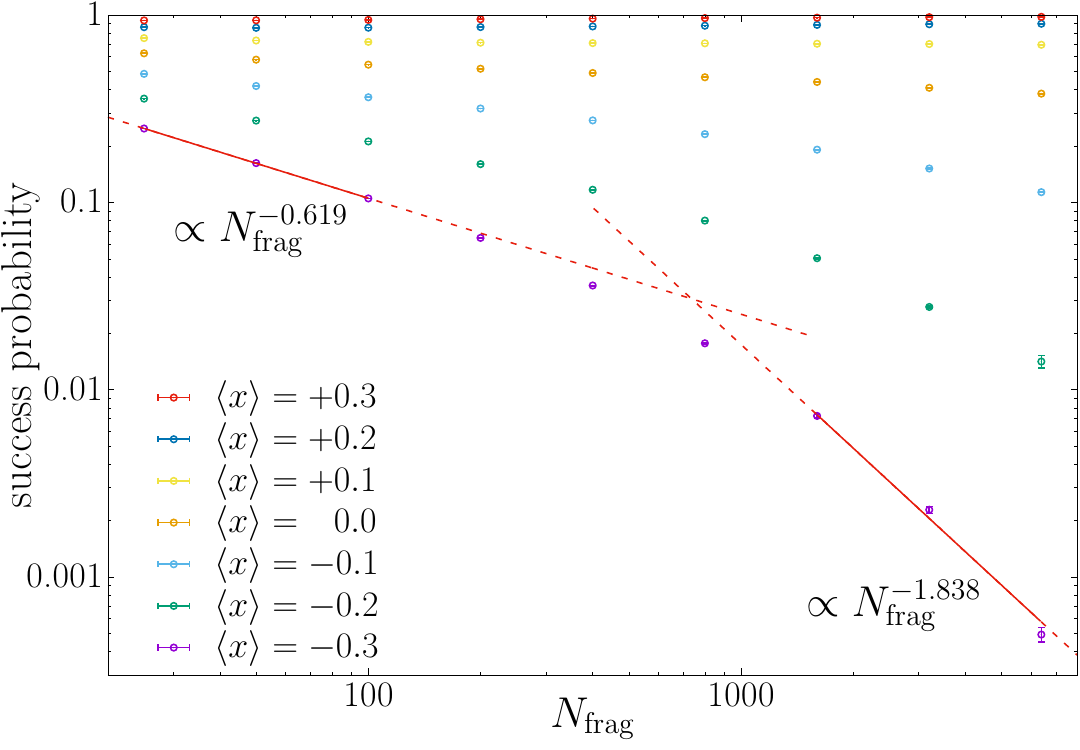} 
\caption{\textbf{The success probability goes to zero in the hard phase.}
If the behaviour shown in Fig.~\ref{fig:gloton-velvet} corresponds
to a phase transition, \ps\ should tend to zero as $\NF\to\infty$ for $\langle x\rangle < \xc$. 
This figure shows that indeed 
\ps\ decays at least as fast as power law in $1/\NF$ in the hard phase, while
in the easy phase our results are already compatible with $\ps=1$ 
for finite sizes.} \label{fig:scaling} \end{figure}

Finally, notice that, in order to have a real phase transition, \ps\ should not
just be very small, but actually go to zero in the hard phase as \NF\ increases.
With our numerical data, see Figure~\ref{fig:scaling}, we can see that the
results are compatible with a power-law decay. On the other hand, for $\langle x\rangle$
significantly larger than \xc, the success probability is compatible with $1$
even for finite \NF.

\begin{figure}[t] 
\includegraphics[width=1\linewidth]{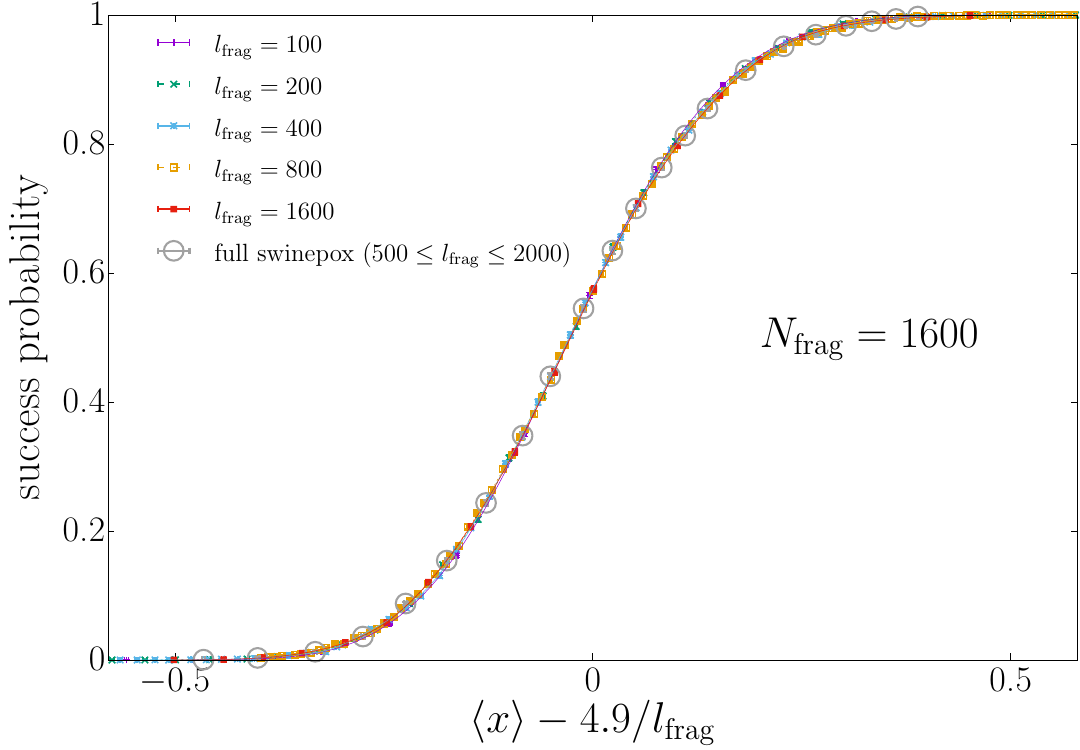} 
\caption{\textbf{Varying the fragment length hardly makes any difference}.   
Our previous results have always considered $\lf=100$. It turns out
that the dependence in this parameter is residual and rapidly 
vanishes as \lf\ grows, according to eq.~(\ref{eq:lf}). That is, the 
curves of \ps as a function of $\langle x\rangle$ can be collapsed
if we subtract the scaling correction caused by finite \lf. 
We also show that the results for a natural
genome (namely that of the swinepox virus) are indistinguishable
from those for random sequences.  In this case, since $L$ is fixed, 
we have a single value of \ps\ for each \lf, all of which
fall on the rescaled curve.
\label{fig:scaling-lfrag}}
\end{figure}
\subsection{The role of the fragment length and comparison with a natural genome}\label{sec:lfrag} 
Thus far, we have always used $\lf=100$ but, given a number of
fragments \NF, two parameters remain: the
genome length $L$ and $\lf$ or, equivalently, $\langle x\rangle$ and
$\lf$.  The beauty of the choice of variables $\langle x\rangle$ and
$\lf$ is that the $\lf$ dependence is residual and vanishes quite fast
as $\lf$ grows, as can be seen in Figure~\ref{fig:scaling-lfrag}. More
precisely,
\begin{equation}\label{eq:lf}
\ps\simeq f[\langle x\rangle + A \log \NF/(\NF \lf)],
\end{equation}
where $A$ is an algorithm-dependent constant\footnote{When relating to
  the continuum results of Ref.~\cite{schlemm:14}, our formulation has
  discretization errors of order $1/L$. The other natural lengths
  ocurring in the problem are $d_{\text{max}}$ and $\lf$, hence we expect
  discretization errors to be controlled by $d_{\text{max}}/\lf$ and
  $d_{\text{max}}/L$. But $d_{\text{max}}/\lf=1-x$ and
  \protect{$d_{\text{max}}/L \sim \log \NF/(\NF \lf)$}, recall
  Eq.~\eqref{eq:discretization}.}. That is, \lf\ acts as a scaling
  correction.

Taking our reads from a random genome or from a
real one seems to make no difference. Indeed, our results for the
success probability for the \emph{Glotón} using reads
sampled from the genome of the swinepox virus
nicely fall onto the same scaled curve obtained for the random genome.
\begin{figure*}[t] \includegraphics[width=.4\linewidth]{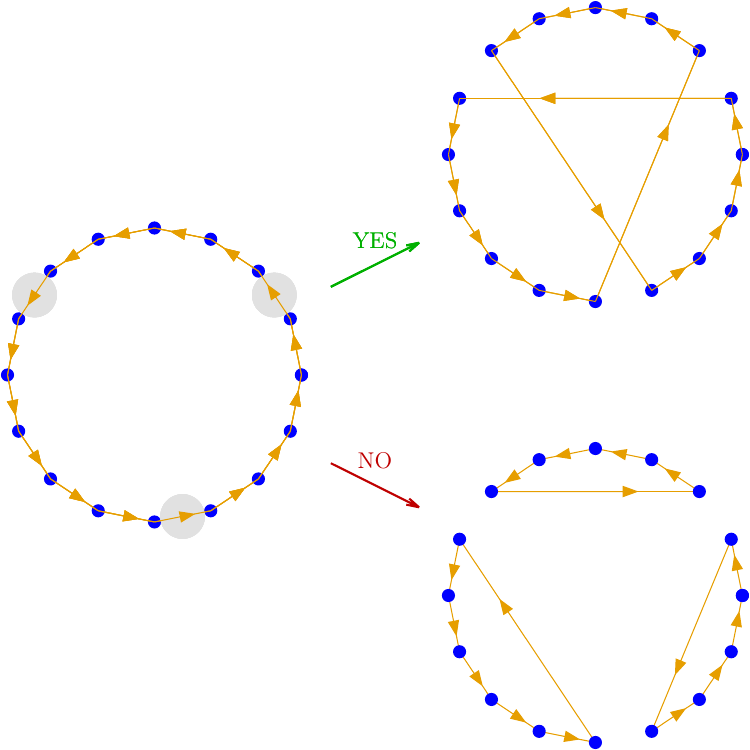}
\includegraphics[width=.58\linewidth]{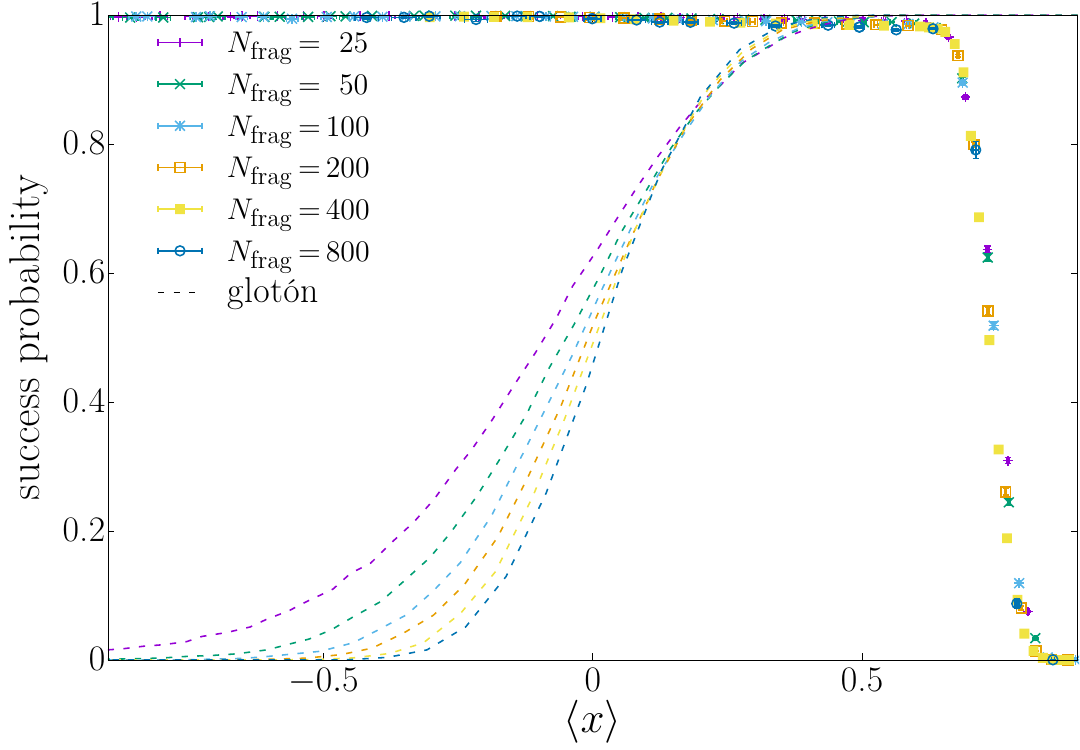}
 \caption{\textbf{A
    Monte Carlo algorithm for the hard cases}. We propose a
  segment-swap Markov-chain Monte Carlo algorithm (sketched in the
  \emph{left} panel) that outperforms common deterministic methods in
  the hard regime (see \emph{right} panel). 
  We represent the permutation of the reads as an
  ordered sequence of fragments (the arrows indicate the sense in
  which the sequence should be toured). The elementary move of the
  algorithm is composed of the following three steps. First, chose
  randomly three independent pairs of \emph{consecutive} fragments
  (depicted with grey circles in the plot), $\ldots \to \alpha_1 \to
  \alpha_2\to\ldots\to \beta_1 \to \beta_2\to\ldots\to \gamma_1 \to
  \gamma_2\to\ldots$. Mind the pair ordering: when one tours the
  circular sequence starting from fragment $\alpha_1$, fragments
  $\gamma_1$ and $\gamma_2$ are not found earlier than $\beta_1$
  and $\beta_2$ (the choices $\alpha_2=\beta_1$ and/or
  $\beta_2=\gamma_1$ are acceptable). Second, consider the rewired
  sequence $\ldots \to \alpha_1 \to \beta_2\to\ldots\to \gamma_1 \to
  \alpha_2\to\ldots\to \beta_1 \to \gamma_2\to\ldots$ (there is an
  unacceptable reconnection ---indicated by \emph{NO} in the figure---
  that would split the sequence into three disconnected cycles). Third
  step: If the new cycle is not longer than
  the original one, the segment swap is
  accepted. As we show in the right panel, segment swap
  is more effective than other algorithms for $-1 < x <
  0.5$.  For $x<0$ the SCS problem no longer corresponds to a full assembly, since
  there are gaps between reads. The segment-swap algorithm, however,
  always finds superstrings that satisfy 
  our success criterion ($\ell \leq
  \ell_\text{ordered}$).  \label{fig:gloton-T0}}
\end{figure*}
  
\section{A better algorithm for the hard phase: the segment-swap}\label{sec:swap}
We have seen that, while
the SCS problem is NP-hard, it becomes solvable for polynomial-time algorithms
in the large-$\langle x\rangle$ regime. For $\langle x\rangle<\xc$, however,
common methods always fail even to find a good approximation to the SCS (provided
in our success criterion by \lo). For moderately negative values of $x$ we have
found a Markov-chain Monte Carlo algorithm that is both powerful and relatively
simple.

The method is sketched in Figure~\ref{fig:gloton-T0}-left. The segment
ordering is actually a circular sequence where we randomly choose
three cutting points.  There are two ways of reconnecting the three
resulting fragments, one of which generates several cycles and can be
discarded. The other reconnection generates a single cycle and is
potentially a new permutation of the reads that effects non-local
changes (one would need
$\sim N_{\text{frag}}$ transpositions of neighboring reads
in order to generate a single segment-swap move).  We accept the new
configuration only if its total length is not larger than in the
previous step. This is the acceptance criterion of a Metropolis
algorithm at zero temperature [see, \emph{e.g.},~\cite{sokal:97}]. As for the
stopping condition, note that there $N_\text{triplets}=\NF!/( (\NF-3)!\,
3!)$ possible choices for the cutting points. Whenever the
length of the sequence has not decreased for 2$N_\text{triplets}$ consecutive
iterations (where the cutting points are chosen randomly with uniform
probability), we check explicitly that none of the
$N_\text{triplets}$ possible moves would decrease the total length. If this is
the case, the run is stopped. 

In spite of its simplicity, this segment-swap method is very
successful in the $-1<\langle x\rangle<0.5$ region and, in particular,
in the negative $\langle x \rangle$ region where both \emph{Glotón}
and \emph{Velvet} fail. The segment-swap method can be generalised by
including a fictive temperature~\cite{kirkpatrick:83} and parallel
tempering~\cite{hukushima:96}. In this way, one would have a candidate
algorithm for treating the $x\to -\infty$ limit of completely
independent reads.  Notice that, as $x$ grows more negative, our
variational solution \lo\ grows worse as an upper bound on the length
of the actual SCS. In these cases, the segment-swap algorithm finds
solutions with $\ell<\lo$, but, at least in the simple $T=0$ version
shown in Figure~\ref{fig:gloton-T0}, we cannot be sure that these
solutions are the actual SCS.

The reader could worry about completeness: is the segment-swap method
capable of reaching all possible configurations? In fact, the
transposition of consecutive fragments is a particular case of the
segment-swap move.  Indeed, take a subsequence
$\ldots\to A \to B \to C\to D\to\ldots$ and let us imagine that one
randomly selects the pairs $A\to B $, $B\to C$ and $C\to D$ as the
ones to be reconnected. In the language of Figure~\ref{fig:gloton-T0},
one would say $\alpha_1=A$, $\alpha_2=\beta_1=B$, $\beta_2=\gamma_1=C$
and $\gamma_2=D$. With this choice, the segment swap results in the
transposition of fragments $B$ and $C$:
$\ldots\to A \to C \to B\to D\to\ldots$. Now, since an arbitrary
permutation may be obtained from an ordered sequence of transpositions
of consecutive fragments, a finite-temperature version of the
segment-swap method is ergodic. Our zero-temperature version of the
algorithm never accepts a move that increases the total sequence
length but, as we said above, this lack of ergodicity causes no
problem in the region $-1<\langle x\rangle<0.5$ (see
Figure~\ref{fig:gloton-T0}). \changes{A plausible explanation is that
  the lack of ergodicity of the zero-temperature dynamics induces
  \emph{another} algorithmic phase transition located near
  $\langle x\rangle=0.5$.}

Another technical question regards the best data structure for
implementing the segment-swap algorithm. We have chosen a linked list,
because the number of operations needed to change the configuration is
independent of $N_{\text{frag}}$. A major drawback, however, is
that one basically needs to go through the full linked list in order
to asses which one of the two possible fragment reconnections is
acceptable. Our solution is
checking the resulting length from both reconnections, \emph{before}
going through the list. Indeed, in most cases, both reconnections
would enlarge the total length and can be rejected without 
checking the list, which requires ${\cal O} (N_{\text{frag}})$
operations. Nevertheless, for larger $N_{\text{frag}}$ than we have
considered in this work, it might be advisable to implement the
segment-swap algorithm with binary-search trees, which have the
potential of turning the computational cost down to ${\cal O} (\log
N_{\text{frag}})$ operations.

\section{Conclusion}\label{sect:conclusion}
We have applied the methods of statistical mechanics in order to
characterise the computational complexity of the SCS problem, showing
that, in terms of an appropriate scaling variable $\langle x\rangle$,
a phase diagram can be constructed. For $\langle x\rangle > \xc$ the
problem is in the easy regime, \emph{i.e.}, it is solvable in
polynomial time, while below \xc\ it is exponentially hard. In the
language of statistical physics, an \emph{order parameter} can be
defined using the probability that a polynomial-time algorithm will
find the correct SCS. For a finite system size (in our case set by the
number of reads \NF) this probability will increase continuously as a
function of the scaling variable (which plays the role of variables
like the temperature, magnetic field or pressure in the phase diagrams
of physical systems). As \NF\ grows, the crossover regime grows
narrower until, in the $\NF\to\infty$ limit, one can speak of a
\emph{phase transition}: the computation always succeeds for $\langle
x\rangle > \xc$ (and always fails for $\langle x\rangle < \xc$).  In
this sense, macroscopic physical systems are considered to be in a
`thermodynamic limit', whose behaviour is indistinguishable from that
of an infinite system. Similarly, while this study has considered
small values of \NF\ in order to penetrate into the hard regime and to
show the scaling behaviour, real instances of assembly 
employ such large \NF\ that one can properly talk of
two distinct phases in its computational complexity.

\changes{Provided that the just-mentioned interpretation of our
  numerical results proves to be correct, the behavior described
  above will be universal}~\cite{amit:05}, in that the same scaling
variable will classify instances of the problem into easy or hard, no
matter which polynomial-time algorithm is used \changesbis{\footnote{In other words,
the universality caused by the presence of a phase transition ensures that all 
polynomial algorithms will fail in the hard phase, \emph{i.e.}, that
the problem is not in P in the usual classification of computational 
complexity.}}. The precise location
of the critical point \xc\ is algorithm dependent, but it will be such
that the average maximum distance between reads is of the order of the
length of the reads, $\langle d_\text{max}\rangle \sim \lf$ (an
intuitive result that we have demonstrated by studying two completely
different methods).  Putting all the above considerations together
with the fact that, in modern high-throughput methods, the genome is
heavily oversampled (implying $\langle d_\text{max}\rangle \ll \lf$)
we have our main result: practical instances of the sequence-assembly
problem are deep in the easy phase, that is, always solvable in
polynomial time. We \changes{would have} thus achieved a
characterisation of the parametric complexity of the problem in the
sense proposed in~\cite{bodlaender:95, nagarajan:09}.

\changesbis{The universality of our result should not be taken to mean that all
algorithms are equally good or, more to the point, that sophisticated
methods based on heuristics and de Bruijn
graphs~\cite{miller:10,zhang:11} are not useful. It does mean
that, as pointed out recently in~\cite{medvedev:21}, their
usefulness does not reside in their turning an NP-hard problem into a
polynomial one. Instead, de Bruijn graphs
are useful because of their efficient implementation for very large
datasets and their power for dealing with errors in the reads and long
repeats in the genomes.}

Below \xc, the SCS problem decouples from that of sequence assembly
(since the full genome cannot be reconstructed unambiguously) and
becomes NP-hard.  In this phase, we do not know the real SCS but we
can consider a variational upper bound on its length given by
\lo. This bound will be good close to \xc\ and deviate more and more
from the real solution to the SCS as $\langle x\rangle$ decreases. We have
explored
this regime using a Markov-chain Monte Carlo method that 
we name segment-swap.
We find that, unlike deterministic methods, our segment-swap
algorithm always succeeds in finding solutions with $\ell\leq\lo$ for
$-1\leq \langle x\rangle\leq 0.5$.  As considered in this work, just
as a proof of concept, the segment-swap method is not ergodic, so there
is no assurance that its stopping point corresponds to the actual
SCS. This shortcoming could be cured by coupling segment swaps with
parallel tempering~\cite{hukushima:96}, which, furthermore, provides a
self-consistent way of validating the
solutions~\cite{janus:10,martin-mayor:15}. We thus believe that
segment-swap Monte Carlo may be considered as a candidate to solve general
instances of the SCS and related problems, such as the travelling
salesman.

\subsection*{Acknowledgments}
This work was partially supported by Ministerio de Ciencia, Innovación y
Universidades (Spain), Agencia Estatal de Investigación (AEI, Spain,
10.13039/501100011033), and European Regional Development Fund (ERDF, A way of
making Europe) through Grant PID2022-136374NB-C21. D.~Y. was supported by the
Chan Zuckerberg Biohub. 

\appendix

\section{Our greedy \emph{Glotón\/} algorithm}\label{sec:gloton}
Our greedy algorithm solves the SCS problem (very
slightly) better than the standard greedy algorithm, of which 
it is only a slight variation [see,
  \emph{e.g.} Ref.~\cite{greedy:greedy}].
As explained in the main text, we represent the superstring 
as a permutation of the original reads (our state space is a
sequence of reads which are consecutive in the permutation, see
Figure~\ref{fig:gloton-T0}--left). The total length of the superstring
is $\NF\lf$ minus the total sum of the overlaps between consecutive
reads (therefore, the SCS corresponds to the maximum
total overlap between consecutive reads).  Our \emph{Glotón\/} seeks the shortest
superstring through the following procedure:
\begin{enumerate}
\item Pick at random the starting fragment of the cycle. This fragment is
  named the \emph{active} read.
\item Consider the overlap with the active read of all the
  still unsorted reads (the \emph{candidates}).
\item Select the candidate that has the maximum overlap with the
  active read.  If there is more than one choice, we pick randomly
  (with uniform probability) one candidate of maximum overlap. The
  chosen candidate is placed in the cycle right after the active read
  and is declared to be the \emph{new} active read.
\item While there are remaining candidates, go back to step 2.
\end{enumerate}
There are two differences with the standard greedy algorithm. First,
the randomness in step 3 (some implementations
pick the maximum-overlap candidate deterministically)
Second, we grow the sequence from just one
active fragment (instead, the greedy algorithm allows more than one
sequence-growing point).

The most demanding part of the \emph{Glotón} algorithm
is the computation of the overlap between fragments. We have sped-up
this part of the computation by generating a look-up table containing
all possible overlaps ---there are $\NF(\NF-1)$
possible ordered pairs of reads. This is particularly useful, because
we run the \emph{Glotón} algorithm $N_\text{attempts}=100$ times
for each given set of $\NF$ reads.

\section{A note on \emph{Velvet}}\label{sec:velvet}
 \emph{Velvet} is not properly
an algorithm for finding the SCS, but instead outputs contigs.  These
are contiguous segments that can unambiguously be inferred to be part
of the original genome. In this case, a successful solution produces a
single contig of length \lo\, while an `unsuccessful' one might be
missing one or more reads or be broken into several contigs. Notice
that this is the program working as desired: it interprets that it
does not have enough data to reconstruct the full genome and, rather
than attempting to find an approximation to an SCS that would not
match the original sequence, it produces unambiguous
subsequences. Hence, the phase transition acquires a different meaning
for Velvet: the critical point separates the region where \emph{the
  fragments database comes from a single contig, for sure,\/} from the
region when Velvet reaches the conclusion that \emph{most probably, the database
  comes from two (or more) contigs,\/} (which is unjustified whenever $x>0$).

\bibliographystyle{apsrev4_1-titles}

\end{document}